\documentclass[letterpaper]{article}
\usepackage{natbib,alifeconf}
\usepackage[latin1]{inputenc}
\usepackage{graphicx,color,longtable,multirow,times,amsmath,url}

\title{FIFA World Cup 2010: A Network Analysis of the Champion Team Play}
\author{C. Cotta$^1$, A.M. Mora$^2$, Cecilia Merelo-Molina \and J.J. Merelo$^2$ \\
\mbox{}\\
$^1$Department of Languages and Computer Science, University of Málaga (Spain) \\
$^2$Department of Architecture and Computer Technology, University of Granada (Spain) \\
ccottap@lcc.uma.es, \{amorag,jmerelo\}@geneura.ugr.es}

\begin{document}
\maketitle

\begin{abstract}
  We analyze the pass network among the players of the Spanish team
  (the world champion in the FIFA World Cup 2010), with the objective
  of explaining the results obtained from the behavior at the complex
  network level. The team is considered a network with players as
  nodes and passes as (directed) edges, and a temporal analysis of the
  resulting passes network is done, looking at the number of passes,
  length of the chain of passes, and the centrality of players in the
  turf. Results of the last three matches
  indicate that the clustering coefficient of the pass network
  increases with time, and stays high, indicating possession by Spanish
  players, which eventually leads to victory, even as the density of
  the pass network decreases with time.
\end{abstract}

%%%%%%%%%%%%%%%%%%%%%%%%%%%%%%%%%%%%%%%%%%%%%%%%%%%%%%%%%%%%%%%%%%%%%%%%%%
%  INTRODUCTION
%%%%%%%%%%%%%%%%%%%%%%%%%%%%%%%%%%%%%%%%%%%%%%%%%%%%%%%%%%%%%%%%%%%%%%%%%%

\section{Introduction}
\label{intro}

The hypothesis that a complex network analysis can help understand
soccer matches has been present for a long time. Several teams, formal
or informally, have performed analysis of soccer matches from the
point of view of the network of passes formed along the match. One
starting point happened in 2004, when a competition to predict the
four best-classified teams in the EuroCup that was celebrated that
year was done in the \emph{Redes} (Spanish for ``network'') social-network mailing list (in
Spanish). The results, which completely failed to predict the outcome,
were published in
\url{http://revista-redes.rediris.es/webredes/eurocopa2004.htm}.

The main problem with these predictions, besides the outcome, which
was completely wrong (not even the two finalists, Portugal and Greece,
were included in any of them) was that they were looking at the static
picture of the team as it emerged from a {\em previous} match. It is
quite clear that soccer is a game of two players, whose networks
clash. While there must be some quantity that is kept from a game to
the next, and that quantity is related to the team structure
(midfielders, forwards, defense), the other team will do its best to
prevent that network to move information (that is, the football) from
one part to other, resulting in a quite unpredictable result.

On the other hand, the network does not have any kind of spatiotemporal information. A network might show a perfect structure, well
formed, with short distances from goalie to forward players, but if it
plays out of place or simply in its own field it will not be able to
obtain a good result. On the other hand, if the network develops quite
slowly with a low number of passes and low precision (low
transitivity), the result will not be good either. In both cases, the
static structure, while meaningful and a good qualitative description
of the overall game, is unable to reflect them.

In this paper, spurred by the victory of the Spanish selection in the
2010 world cup, we have performed a spatiotemporal analysis of the
essential games that led to victory. In this analysis, we have looked
at the temporal evolution of the number of passes, the centrality of
the player and pitch zones networks, and also at the length of the
chains of passes and its transitivity (taking thus into account the
effect of the opponent of simple ball losses). In this way, we take
into account both the complex network structure (reflected in the
power-law structure of the length of the number of passes and overall
network structure) and the spatiotemporal nature of the game. In that
sense, this paper is the first to make that kind of analysis, which
can latter be complemented with other kind of static micro, macro, and
meso measurements of the same type.

The rest of the paper is structured as follows: next, we examine the
state of the art in analysis of the outcome of soccer matches. Then,
the methodology used to extract data from the match is presented. An
overall examination of the matches
played by the Spanish team
is subsequently performed. We close the paper with some
conclusions and guidelines for future work.

%%%%%%%%%%%%%%%%%%%%%%%%%%%%%%%%%%%%%%%%%%%%%%%%%%%%%%%%%%%%%%%%%%%%%%%%%%
%  PRELIMINARY CONCEPTS - COMPLEX SYSTEMS IN SPORTS
%%%%%%%%%%%%%%%%%%%%%%%%%%%%%%%%%%%%%%%%%%%%%%%%%%%%%%%%%%%%%%%%%%%%%%%%%%

% \section{Complex Systems in Sports}
% \label{CSSports}

% some preliminaries

%%%%%%%%%%%%%%%%%%%%%%%%%%%%%%%%%%%%%%%%%%%%%%%%%%%%%%%%%%%%%%%%%%%%%%%%%%
%  STATE OF THE ART
%%%%%%%%%%%%%%%%%%%%%%%%%%%%%%%%%%%%%%%%%%%%%%%%%%%%%%%%%%%%%%%%%%%%%%%%%%

\section{State of the Art}
\label{SotA}

Despite the huge cultural and popular interest that soccer arises,
being arguably the most popular sport (or maybe spectacle) in the
world, there have not been many scientific approaches to sport
performance and prediction. This was true when \cite{onody2004complex}
wrote their often-referenced paper in
2004, and it is still true today.

However, since then, several papers try to apply complex network
analysis to the soccer world. The above-mentioned paper itself was
seminal in its thoroughness: it analyzed the network of all Brazilian
soccer players, and linked them if they had shared a team, and found
that several metrics (number of teams by player, number of goals by
player, number of games played) follow truncated power laws or
exponential distributions. However, there was no attempt to relate any
of those quantities to performance. Could the number of teams a player
has participated be related to the number of goals?

In general, the prediction of performance has concentrated on the
hypothesis that a team has some kind of intrinsic fitness --see
\cite{springerlink:10.1140/epjb/e2009-00024-8}-- whose difference
between teams affect the probability (not the certainty) of one
beating the other. These studies have mainly concentrated on time
series analysis \citep{heuer2010soccer} but not intra-game dynamics.

Since the pass data for several world-class events (Euro Cup and World
Cup) were made available, one of the authors has been doing informal
analysis on the team's networks and deducting from them some kind of
qualitative prediction on the result of a match. However, this
analysis was not formal at all, and even as differences between the
network qualities of different teams were appreciated, it was
difficult to relate them either to the team fitness or to the match
outcome. The ARSf\'utbol team, based in Argentina, has done extensive
analysis of world-class events as well as local low-level soccer
teams \citep{bundio2009analisis}, concluding that the performance of a
team is mainly related to the existence of a well-coordinated core of
players (such as the players coming from FC Barcelona in the Spanish
national team or the set of players from Porto FC in the Portuguese
selection). Even those post-hoc observations cannot be easily used for
predicting performance.

In this paper, we will look at the micro-dynamics of a soccer team,
the Spanish selection during the latest World Cup in which they
emerged as champions, bringing happiness to millions of Spaniards all
over the world. We will try to find out which quantities made the
Spanish team excellent by focusing on quantitative analysis of its
game-play. This will be used as a first step for a second leg of
analysis which will be focus on prediction.

%%%%%%%%%%%%%%%%%%%%%%%%%%%%%%%%%%%%%%%%%%%%%%%%%%%%%%%%%%%%%%%%%%%%%%%%%%
%  DATA EXTRACTION METHODOLOGY
%%%%%%%%%%%%%%%%%%%%%%%%%%%%%%%%%%%%%%%%%%%%%%%%%%%%%%%%%%%%%%%%%%%%%%%%%%

\section{Data extraction}
\label{data_extract}
\begin{figure}[t!]
\begin{center}
\includegraphics[scale=0.20]{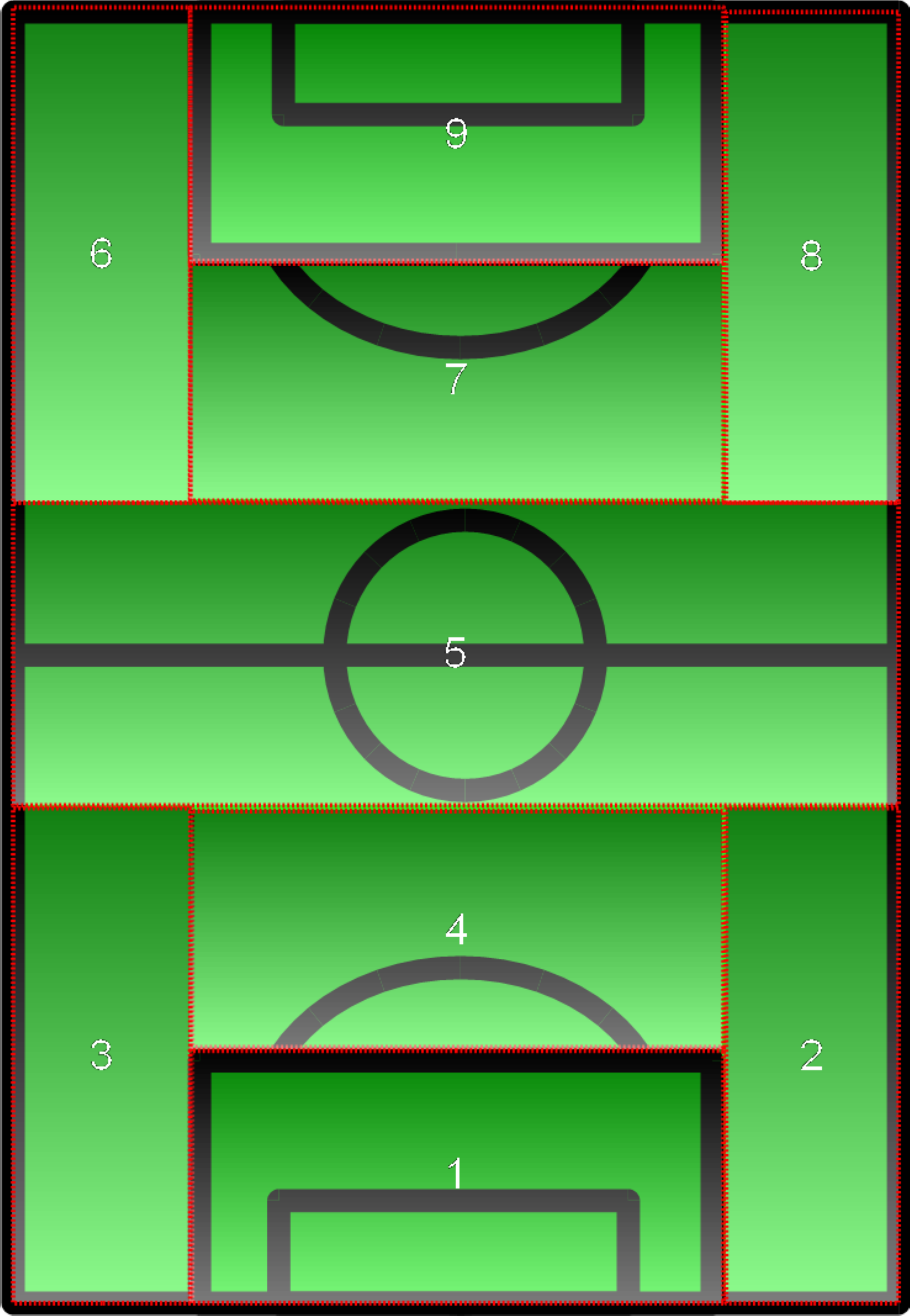}
\caption{Model of the football field for data extraction. Playing zones.
\label{fig:field_model}}
\end{center}
\end{figure}

In order to collect the data for the analysis, an important issue was the fact that we intended to analyze the game play from a spatiotemporal perspective. The spatial dimension tries to capture the fact that a player can act in different parts of the pitch and his role (and therefore the way he performs and interacts with the teammates) may be different in each of these zones. For example, a certain midfielder can have a defensive role when the opposing team is attacking, and have the teammates in the last defensive line as preferred target for a pass whenever he regains control of the ball. On the other hand, this very same player can adhere to a much more combinative play when in offensive positions, interacting more with wingers and other creative midfielders. To this end, the pitch has been divided into nine zones as depicted in Figure \ref{fig:field_model}: four zones correspond to defensive positions (own box, wingback lanes, and own midfield), other four zones correspond to offensive positions (opponent's box and midfield and wingers' lanes), and the the very central zone of the pitch which plays an important role for both initiating static attacks and pressing the opponent in defense. For the purpose of this analysis, each player is figuratively divided into 9 different players depending of the pitch zone in which they touch the ball (of course, most players only visit a limited number of zones throughout a given game).

As to the temporal perspective, our goal is to capture the fact that a
football game can go through different phases in which a team can
change its way of playing (e.g., the dominant team can become
dominated even if just sporadically or the trainer may introduce
tactical changes) or at least can change its effectiveness (e.g., a
player that was nullified by an effective defense can resurface later
on, when tiredness prevents the defenders for keeping on tight
marks). To account for this, we also keep track of the minute in which
each pass was done.

Once the kind of data we needed was defined, data extraction was done
by ourselves, reviewing the knock-out games played by Spain
(contrarily to regular group games in which a team can speculate with
a draw or even with a somehow minimal loss in order to qualify for the
next phase, knock-out games imply a win-loss situation and therefore
the playing style and tactics is effectively directed at winning the
game without depending on external factors\footnote{Winning the game
  must be interpreted here as progressing on to the next knock-out
  stage rather than, e.g., scoring more goals than the opponent. For
  example, a team may decide to play very defensively in order to
  reach the penalty shootout if they consider they will be outplayed
  otherwise; be it as it may, this reflects a definite and valid style,
  and is anyway not applicable to the games considered in which Spain
  went effectively to win the game before penalty shootouts.}). These
correspond to the games against Portugal (1/8th) (not analyzed in this
paper), Paraguay (1/4th), Germany (semifinals) and the Netherlands
(final). The raw data obtained from this visual inspection consists of
a list of passes with the format\vspace{2mm}

\centerline{$<$\emph{half}$>$ $<$\emph{minute}$>$ $<$\emph{player}$>$ $<$\emph{zone}$>$ $<$\emph{player'}$>$ $<$\emph{zone'}$>$\vspace{2mm}}

\noindent indicating which player passed to which player, in which
minute, and in which pitch zones were each of them. As mentioned
before, each pair (\emph{player}, \emph{zone}) can be interpreted as a
virtual player for the purposes of network construction. More
precisely we consider a moving window of 15 minutes (which we believe
is long enough to capture the state of the game at any given instant)
and build a series of directed graphs $G_i(V,E_i)$ where $V=\{(p,z)\
|\ p\in T,\ z\in\{1,\cdots,9\}\}$, $T$ is the set of actual players in
the national team, and $(u\rightarrow v)\in E_i$ if, and only if,
virtual player $u$ passed to virtual player $v$ within the $i$-th time
window.

This data was extracted by visualizing the matches by the authors;
data collection took approximately 4-5 times the duration of
matches\footnote{This data is available to the scientific community,
  with the provision that this paper is referenced should any publication arise from the study.}.

%%%%%%%%%%%%%%%%%%%%%%%%%%%%%%%%%%%%%%%%%%%%%%%%%%%%%%%%%%%%%%%%%%%%%%%%%%
%  SPANISH TEAM ANALYSIS (DATA RESULTS)
%%%%%%%%%%%%%%%%%%%%%%%%%%%%%%%%%%%%%%%%%%%%%%%%%%%%%%%%%%%%%%%%%%%%%%%%%%

\section{The Spanish Team Analysis}
\label{spanish_team_analysis}

The Spanish national team attended the World Cup 2010 as one of the
favorites to win the trophy. While to some extent this was not a
completely new situation from the local point of view (and even the
scientific point of view: Spain hosted the world cup in 1982, and New
Scientist published an article proving why it should win due to home
advantage \citep{dowie1982spain}), in this case this consideration was
quite reasonable a priori, since Spain was the current incumbent in
Europe after winning the Euro Cup in 2008 for the 2nd time in history
(the 1st time dating back to 1964 in the Euro Cup hosted precisely in
Spain). This win at the European level (and even more precisely the
win over Italy in penalty shootouts in the 1/4th game) was seen by
many as a turning point from the psychological point of view, finally
breaking the ``curse of the 1/4ths'' that had been unescapable for
several other --very talented-- Spanish teams. On reflection, the
current generation of players was probably the most adequately suited
for this purpose, since they had been surfing on a winning wave in the
national team since junior stages (\'Iker Casillas, Xavi Hern\'andez
and Carlos Marchena were members of the team than won the U-20 World
cup in 1999) and were also part of the last dominant cycles of Real
Madrid CF and FC Barcelona at club level (holding numerous trophies a
European and International club level).

The lack of this ``choking culture'' that had haunted preceding
generations, and the final discovery of a playing style that players
felt their own (incidentally a style that departed by long from the
traditional ``Spanish fury'' that had been the national team's
trademark until them, and arguably a finding that was in part possible
due to the unfortunate series of injuries of several players) led to
the current success cycle of the the team. While this style
--informally termed ``tiki-taka'' (an otherwise meaningless phrase
that can be translated as ``touchy-touch'') by the late Andr\'es
Montes, a famous TV commentator-- is now considered a ``dogma of
faith'' by many in Spain, and is commonly assimilated to that used by
FC Barcelona (a claim based on the number of players of this club in
the national team), a deeper inspection indicates that there are
definite stylistic differences not only between the national team and
other clubs in Spain, but also between the Euro 2008 team and the
World Cup 2010 team. The former had players such as Fernando Torres
(then at Liverpool FC) or David Silva and David Villa (then at
Valencia CF) at their highest performance level and played a much more
direct, faster combinative football. The latter team had several of
these players at a lower performance level, some key players in
midfield or defense were obviously two years older, and other players
more suited to passing play were incorporated. The result was a slower
playing pace in which the ball and players move rapidly, but in which
the net advance is slow (yet steady). In the following we will analyze
some of the features of this style using the network information
collected as indicated in Section \ref{data_extract}.

% -------------------------------------------------------------------------

\subsection{Quarter Final Match: Paraguay-Spain}
\label{quarter}

\begin{figure}[!t]
\begin{center}
\includegraphics[scale=0.60]{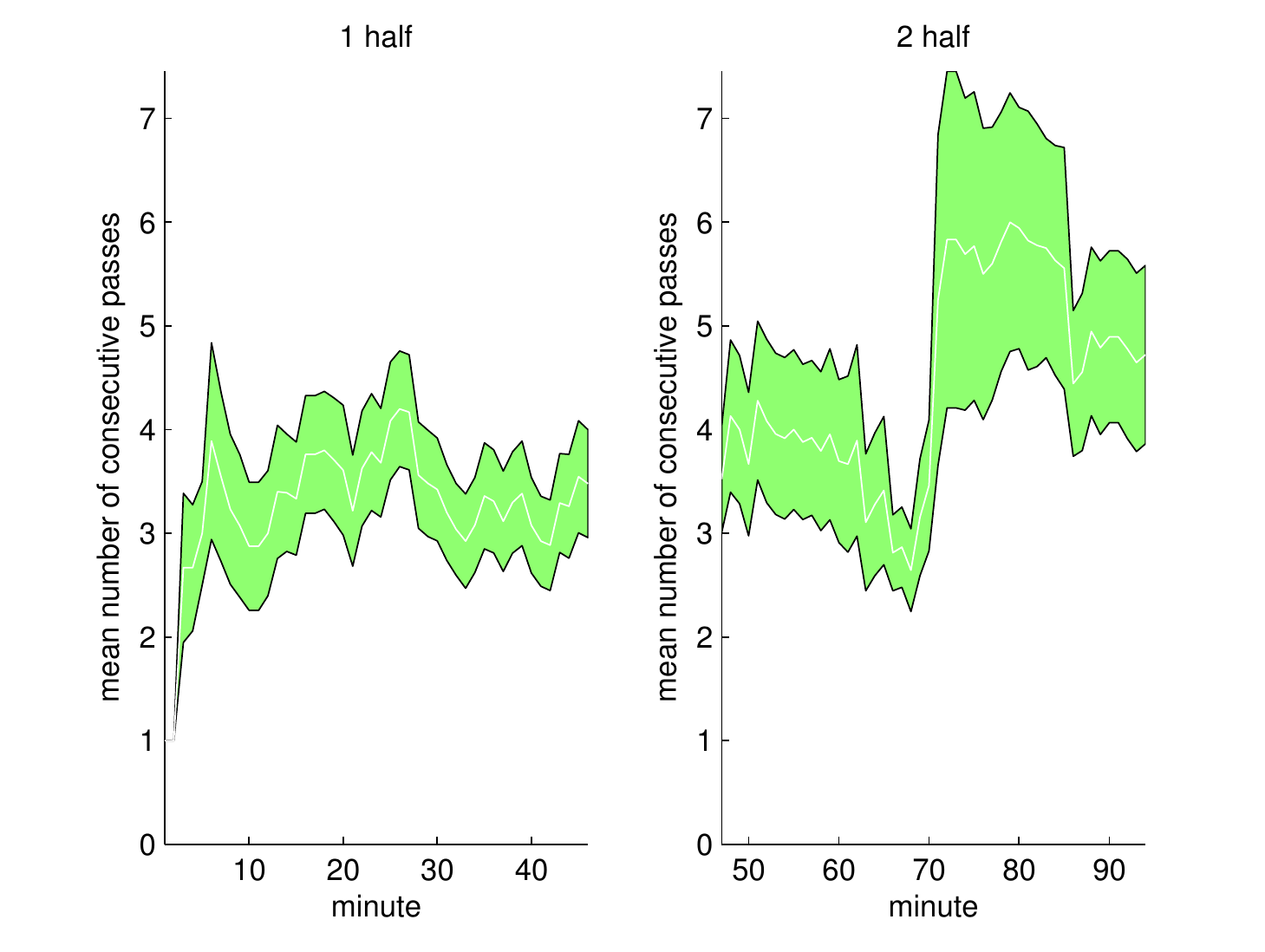}
\caption{Paraguay-Spain. Average number of consecutive passes. In this figure and all subsequent ones, the white line is the mean and the shadowed area covers one standard error of the mean above and below the former.
\label{fig:PvsS-avg_pass_length}}
\end{center}
\end{figure}

\begin{figure}[!ht]
\begin{center}
\includegraphics[scale=0.60]{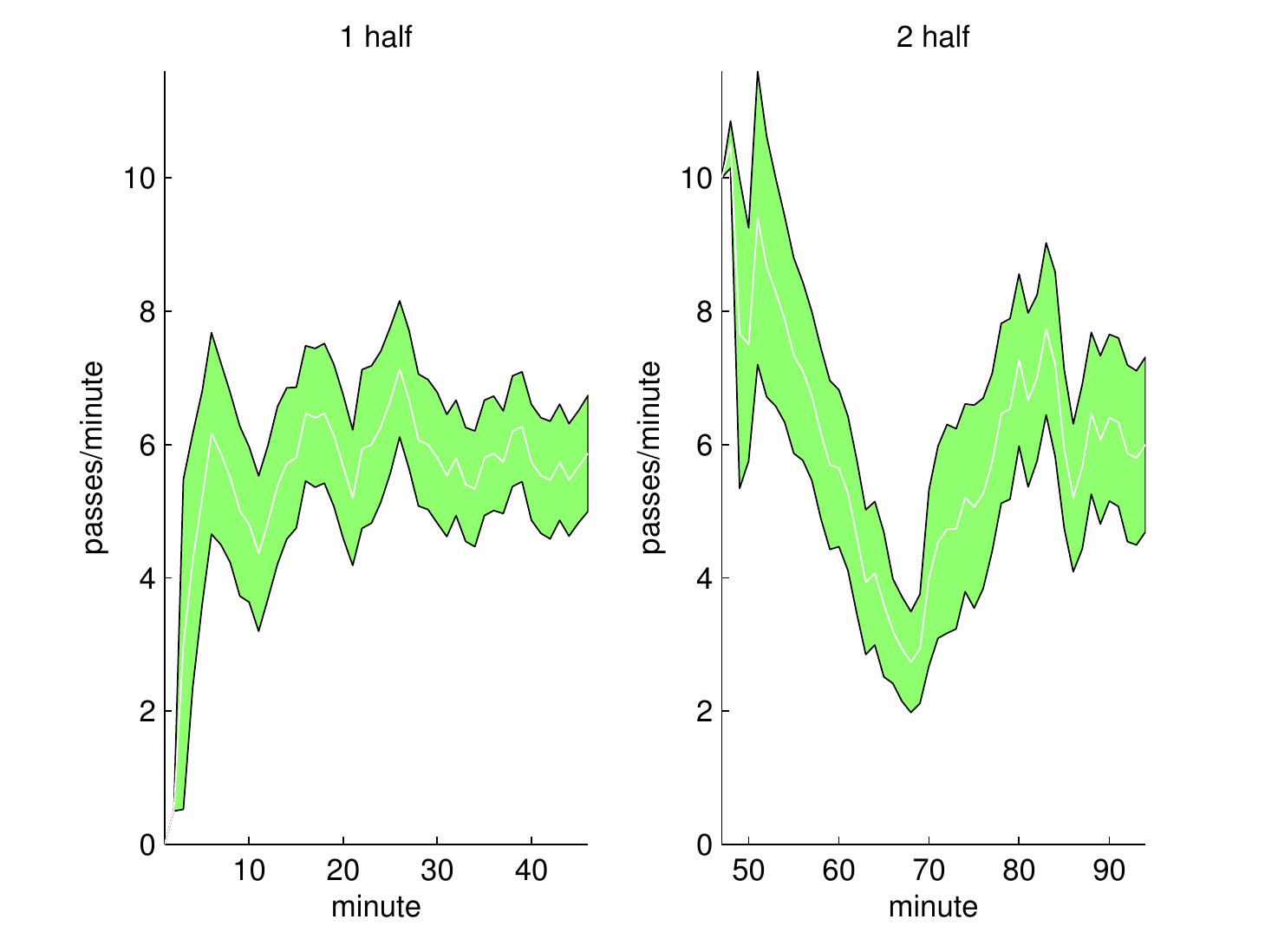}
\caption{Paraguay-Spain. Passes per minute (right).
\label{fig:PvsS-pass_minute}}
\end{center}
\end{figure}

\begin{figure}[!t]
\begin{center}
\includegraphics[scale=0.60]{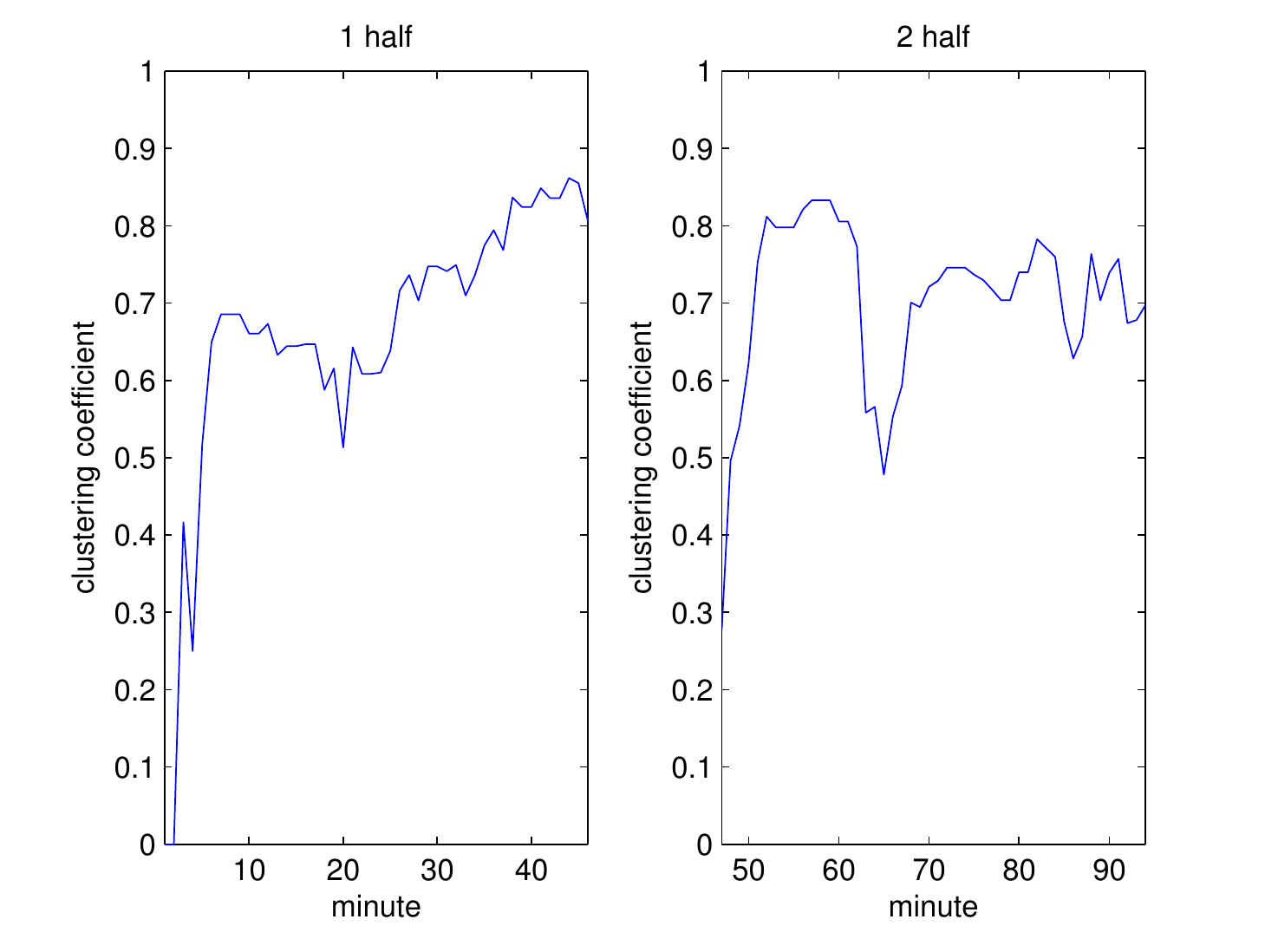}
\caption{Paraguay-Spain. Clustering rate.
\label{fig:PvsS-cluster_rate}}
\end{center}
\end{figure}

\begin{figure*}[!ht]
\begin{center}
\includegraphics[scale=0.5]{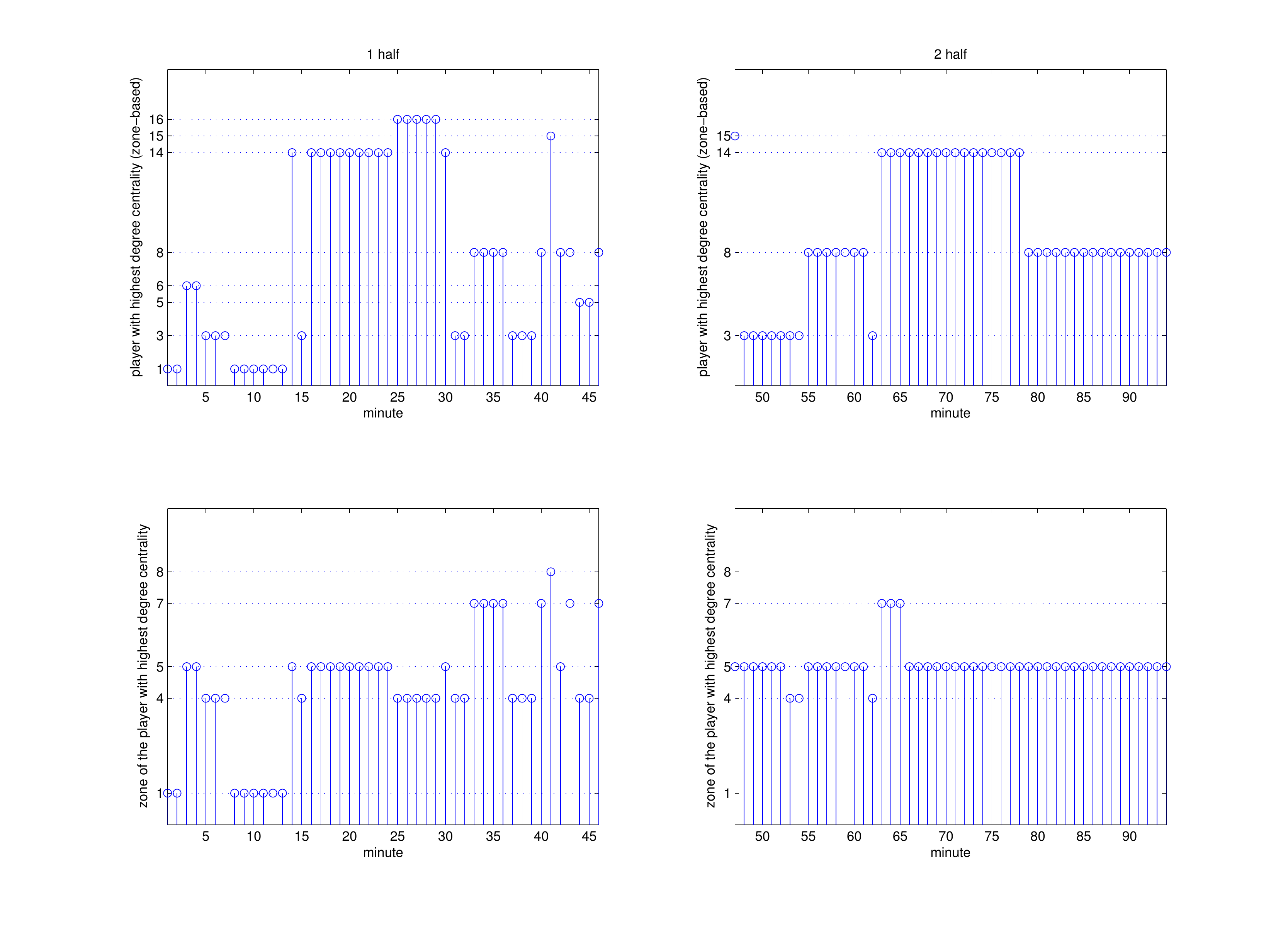}
\caption{Paraguay-Spain. Zone-based player centrality.
\label{fig:PvsS-player_centrality}}
\end{center}
\end{figure*}

Spain reached quarterfinals after winning 1-0 (a result that would
repeat itself in all remaining games) against Portugal. While the
result of that last game was tight (and whether Villa, the scorer of
the sole goal was in off side is open to some question), from a global
point of view Spain was widely regarded as having dominated the game
and deservedly progressing to quarterfinals, the historically
unsurmountable barrier for the national team. The game was approached
by Paraguay (a strong opponent, as demonstrated by their being
runner-ups in the Copa Am\'erica 2011) from a tactical premise: to
disrupt the flow of the ball among Spanish players. They succeeded at
it for most of the game, with Spain hardly achieving more than 3-4
consecutive passes (see Figure \ref{fig:PvsS-avg_pass_length}) and
only attaining a moderate pace of about 6 passes per minute (see
Figure \ref{fig:PvsS-pass_minute}). Although Spain eventually settled
to its usual combinative game (see clustering coefficients in Figure
\ref{fig:PvsS-cluster_rate}), the first half was very irregular for
the team, with no single player emerging as the most central (in terms
of in-/out-degree), and with the game taking place mostly in Spain's
defensive zone (see Figure \ref{fig:PvsS-player_centrality}). This
does not mean Spain was necessarily under attack, but that Paraguay's
pressing avoided Spain's settling the ball in offensive zones. The
game's turning point took place with two penalty shoots failed
successively by Paraguay and Spain (low valley shown in Figure
\ref{fig:PvsS-pass_minute}--right), after which Paraguay lowered their
defensive strength due to tiredness, and Spain could finally enter
into longer sequences of passes, with Xavi Hern\'andez and Xabi Alonso
emerging as dominant players in the midfield. 


% -------------------------------------------------------------------------

\subsection{Semi Final Match: Germany-Spain}
\label{semi}
The victory over Paraguay in the quarterfinals meant Spain advanced for the first time in history to semifinals\footnote{The previous best performance of Spain had been a 4th place in Brazil's World Cup 1950, an edition in which there was a small league among the top-four teams.}, and a general consensus that the team had fulfilled the minimum expectations of the supporters. This meant that the semifinal was approached as a game in which there was nothing to loose (although it must be noted that the same held for Germany, who had presented a very young team and a new, less physical style, and were in a transitional year). Maybe due to this lack of pressure, Spain put forward their best performance in all the tournament.

\begin{figure}[!t]
\begin{center}
\includegraphics[scale=0.60]{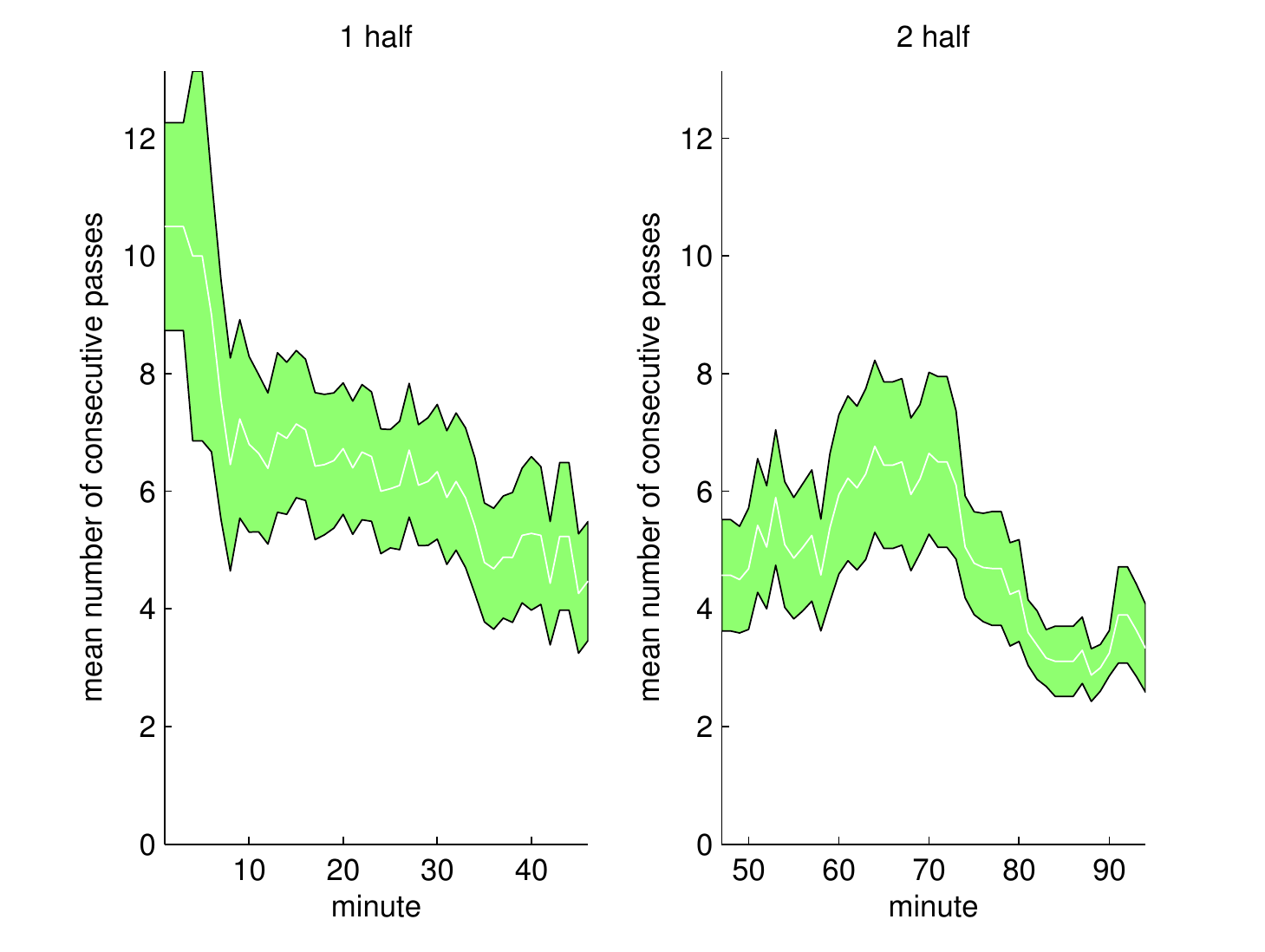}
\caption{Germany-Spain. Average number of consecutive passes.
\label{fig:GvsS-avg_pass_length}}
\end{center}
\end{figure}

\begin{figure}[!ht]
\begin{center}
\includegraphics[scale=0.60]{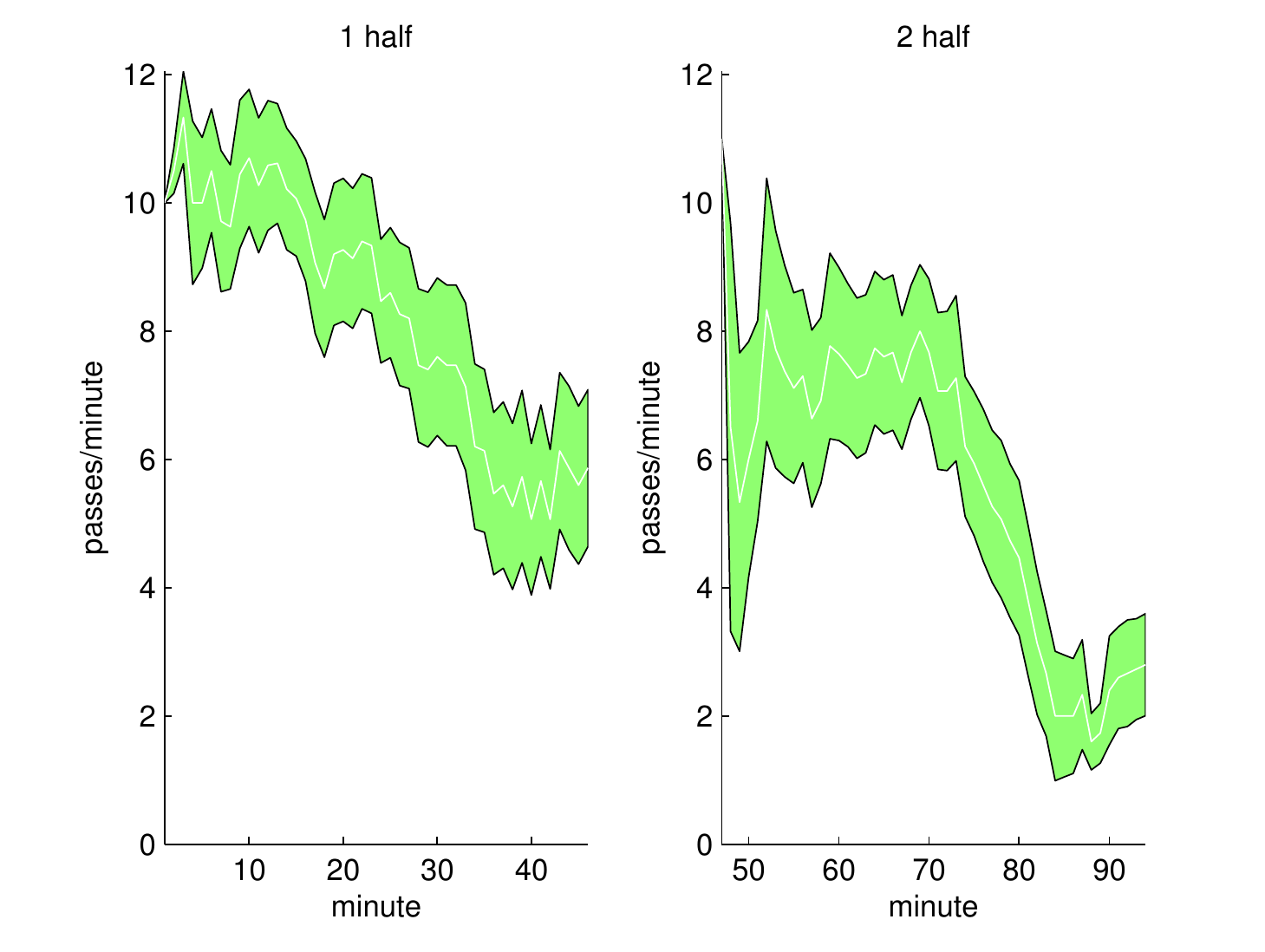}
\caption{Germany-Spain. Passes per minute (right).
\label{fig:GvsS-pass_minute}}
\end{center}
\end{figure}

\begin{figure*}[!ht]
\begin{center}
\includegraphics[scale=0.5]{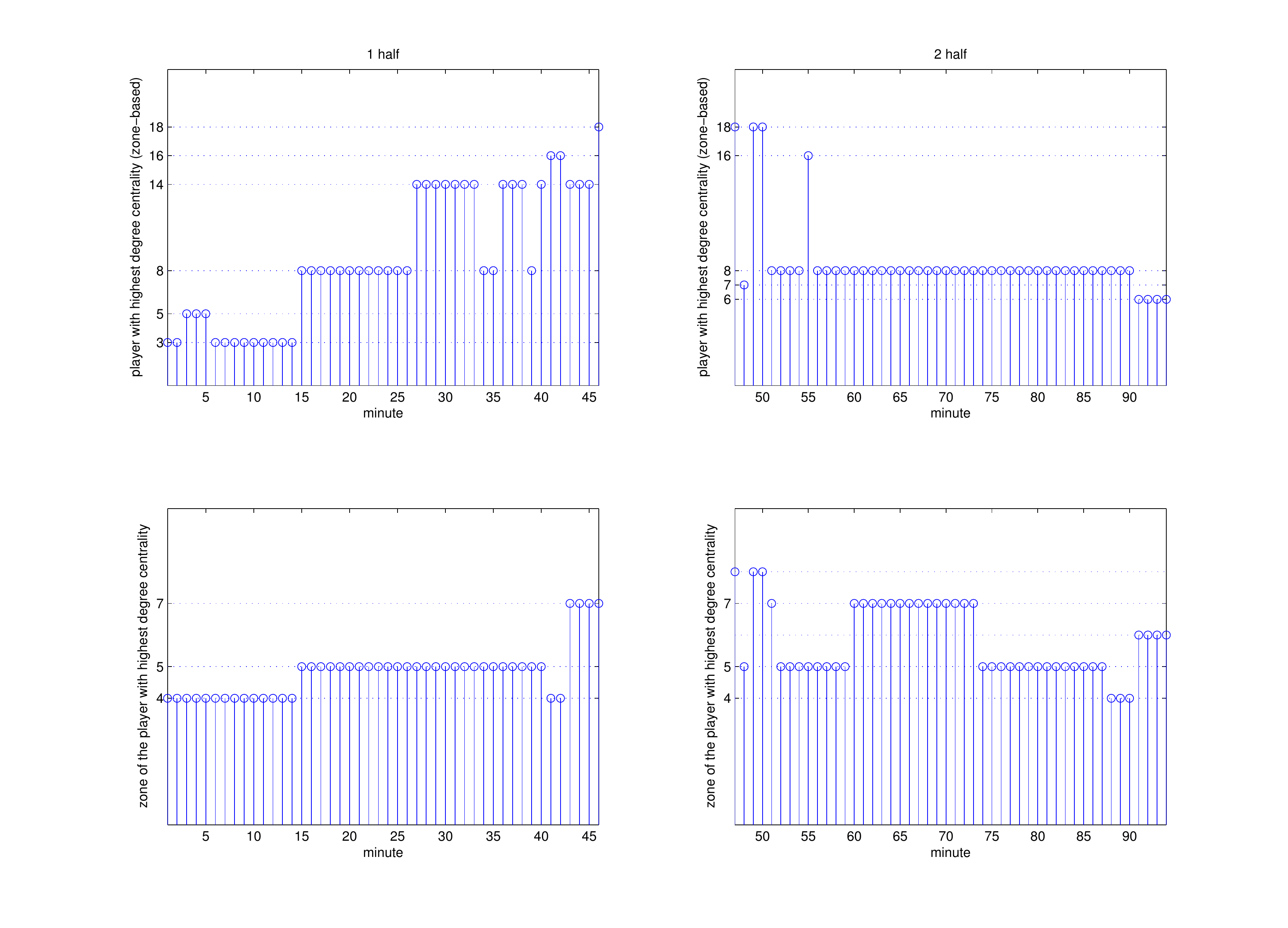}
\caption{Germany-Spain. Zone-based player centrality.
\label{fig:GvsS-player_centrality}}
\end{center}
\end{figure*}

The game was thoroughly dominated by Spain, who managed to engage in a fast game (well above 6 passes per minute, the ceiling reached for the most part of the game against Paraguay, cf. Figures \ref{fig:PvsS-pass_minute} and \ref{fig:GvsS-pass_minute}) and more elaborated sequences of passes (roughly twice as much as against Paraguay, cf. Figures \ref{fig:PvsS-avg_pass_length} and \ref{fig:GvsS-avg_pass_length}). It is interesting to note that the clustering coefficient was rather stable (the drop near the end of the 2nd half corresponds to the final push exerted by Germany once they were 1 goal down) and similar to that reached against Paraguay, which indicates the underlying passing pattern was maintained in both games (but was more effective --i.e., resulted in longer sequences of passes-- against Germany). Xabi Alonso and specially Xavi Hern\'andez during the 2nd half were the clear hubs of the game (Paraguay had partly succeeded in disabling these two players, hence the more disconnected game). Actually the 2nd half saw the ball parked deeply inside the German field (Figure \ref{fig:GvsS-player_centrality}--bottom, right), indicating a clear domination of the game: sequences of up to 24 passes took place before Spain's goal (which paradoxically was a header after a corner kick).

\begin{figure}[!t]
\begin{center}
\includegraphics[scale=0.60]{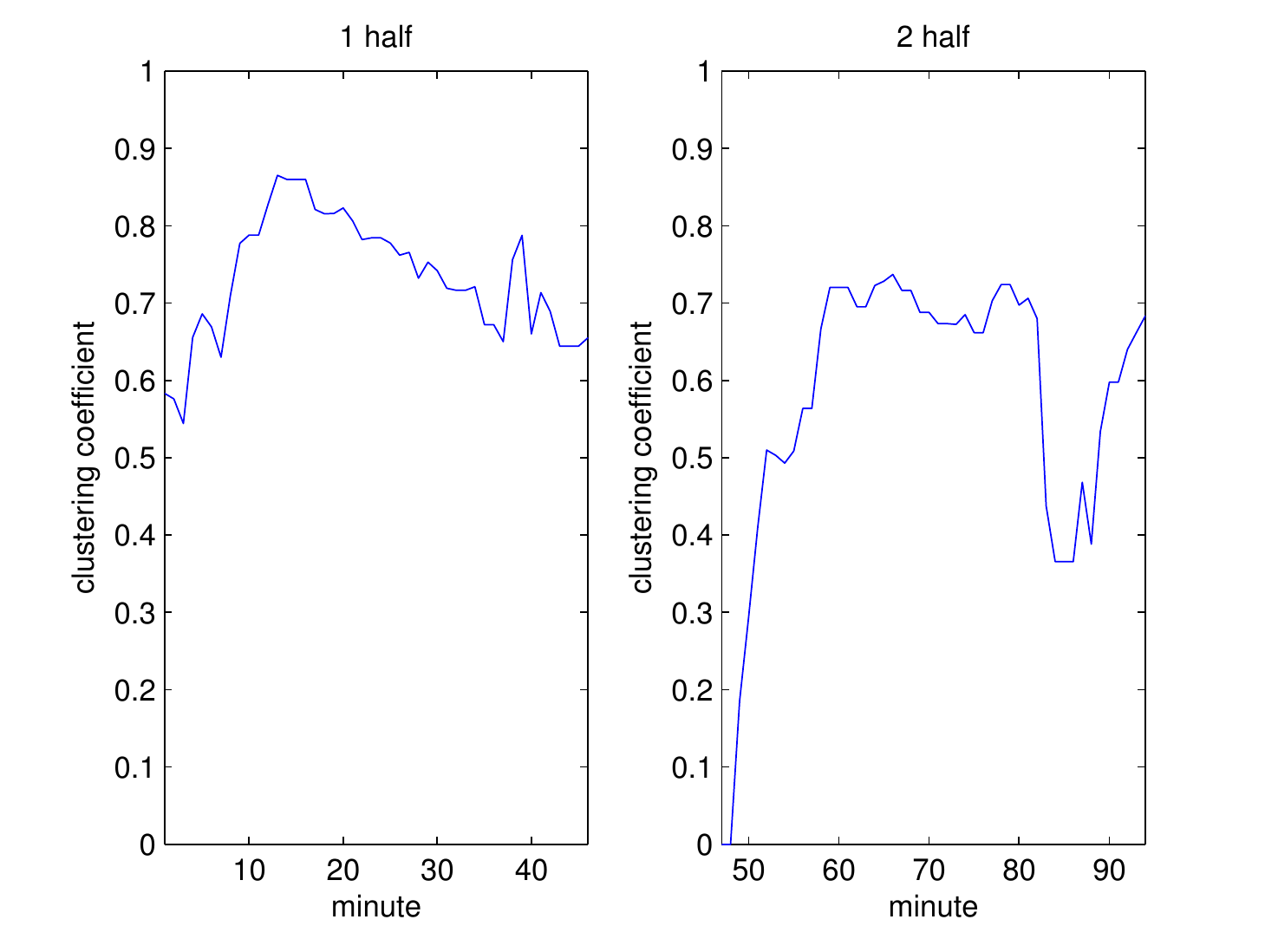}
\caption{Germany-Spain. Clustering rate.
\label{fig:GvsS-cluster_rate}}
\end{center}
\end{figure}
%
%% -------------------------------------------------------------------------
%
\subsection{Final Match: Netherlands-Spain}
\label{final}
\begin{figure}[!ht]
\begin{center}
\includegraphics[scale=0.60]{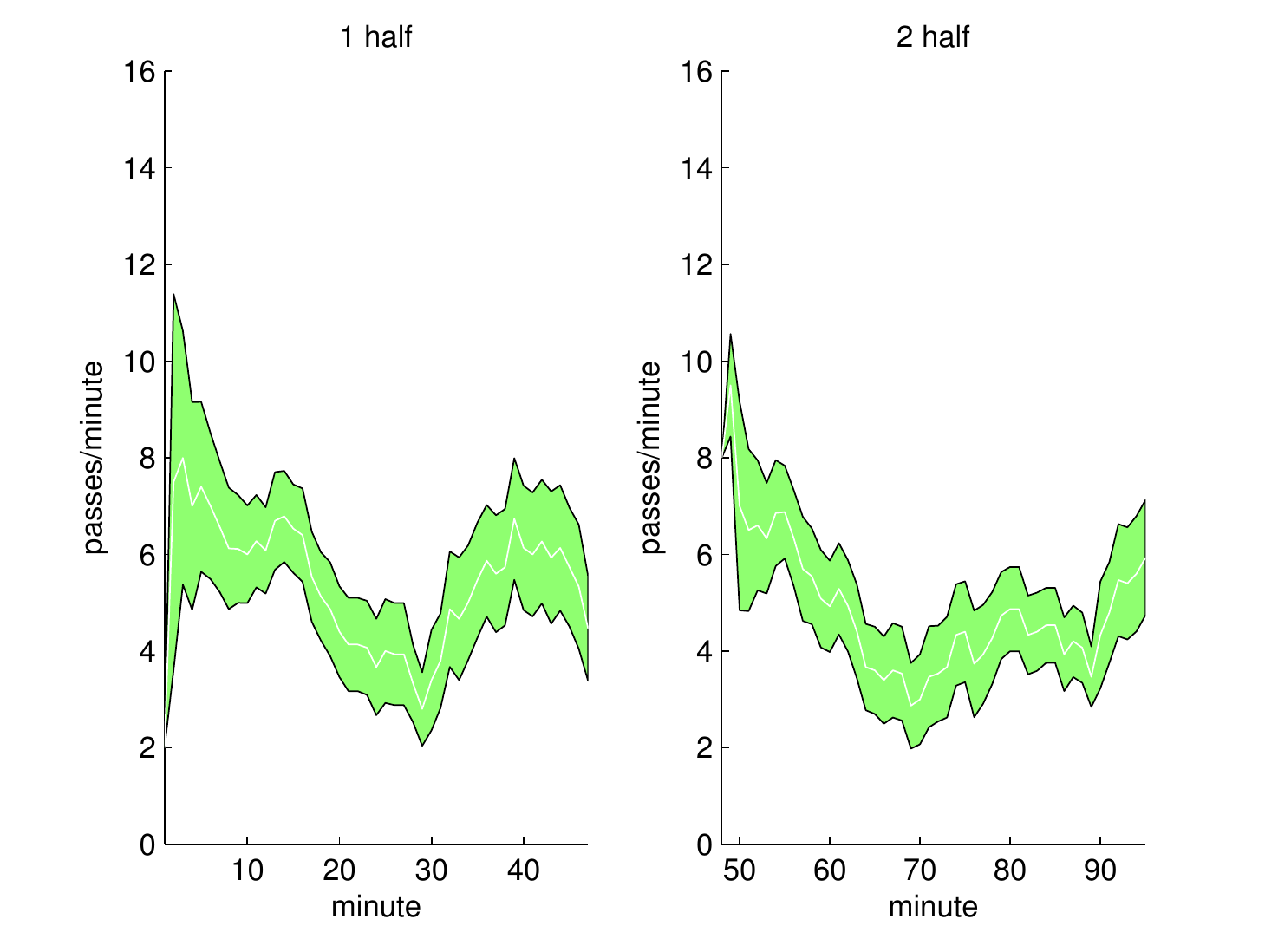}
\caption{Netherlands-Spain. Passes per minute.
\label{fig:HvsS-pass_minute}}
\end{center}
\end{figure}

\begin{figure}[!ht]
\begin{center}
\includegraphics[scale=0.60]{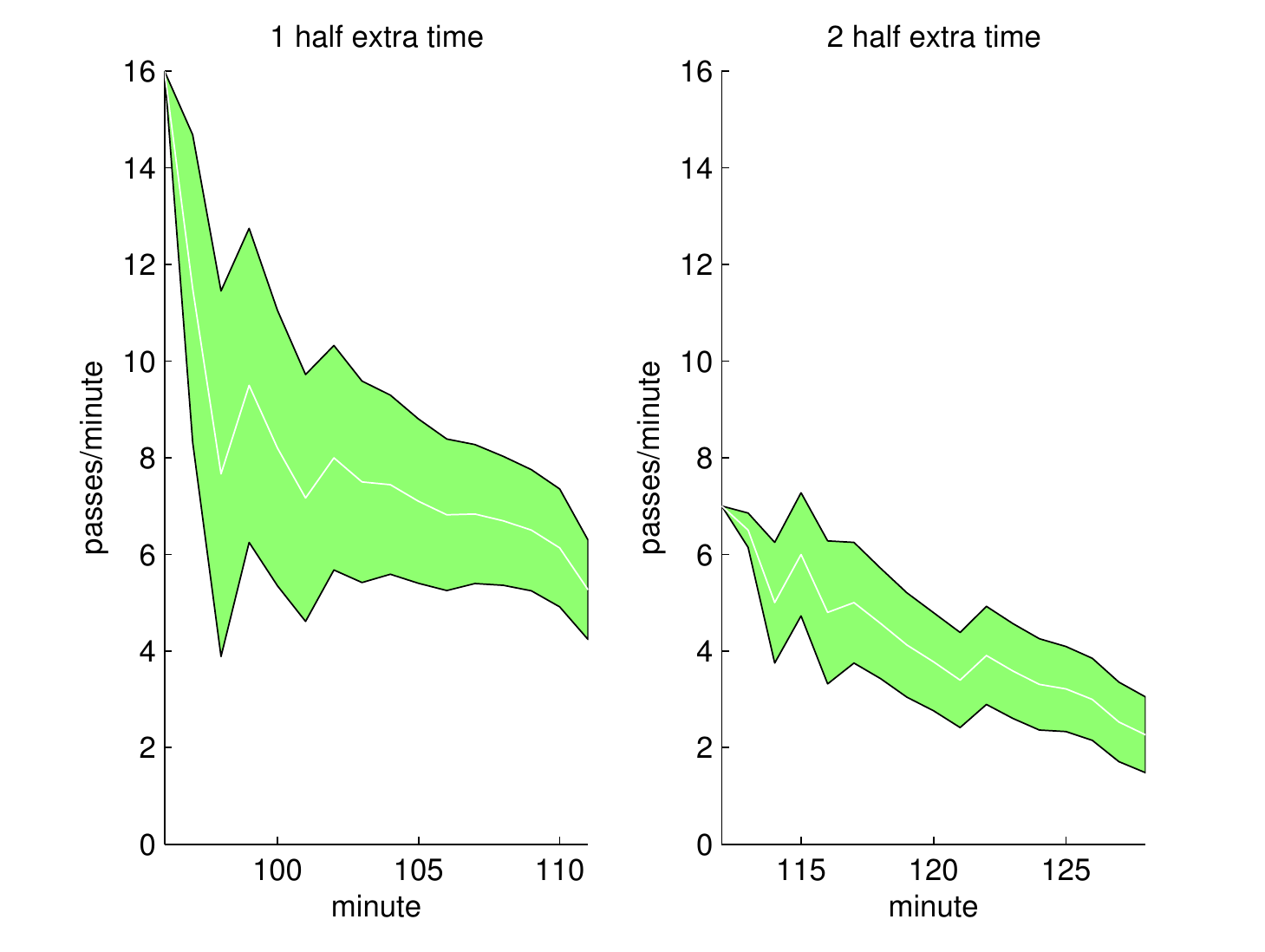}
\caption{Netherlands-Spain. Passes per minute in the extra time.
\label{fig:HvsS-pass_minute-ET}}
\end{center}
\end{figure}

\begin{figure}[!t]
\begin{center}
\includegraphics[scale=0.60]{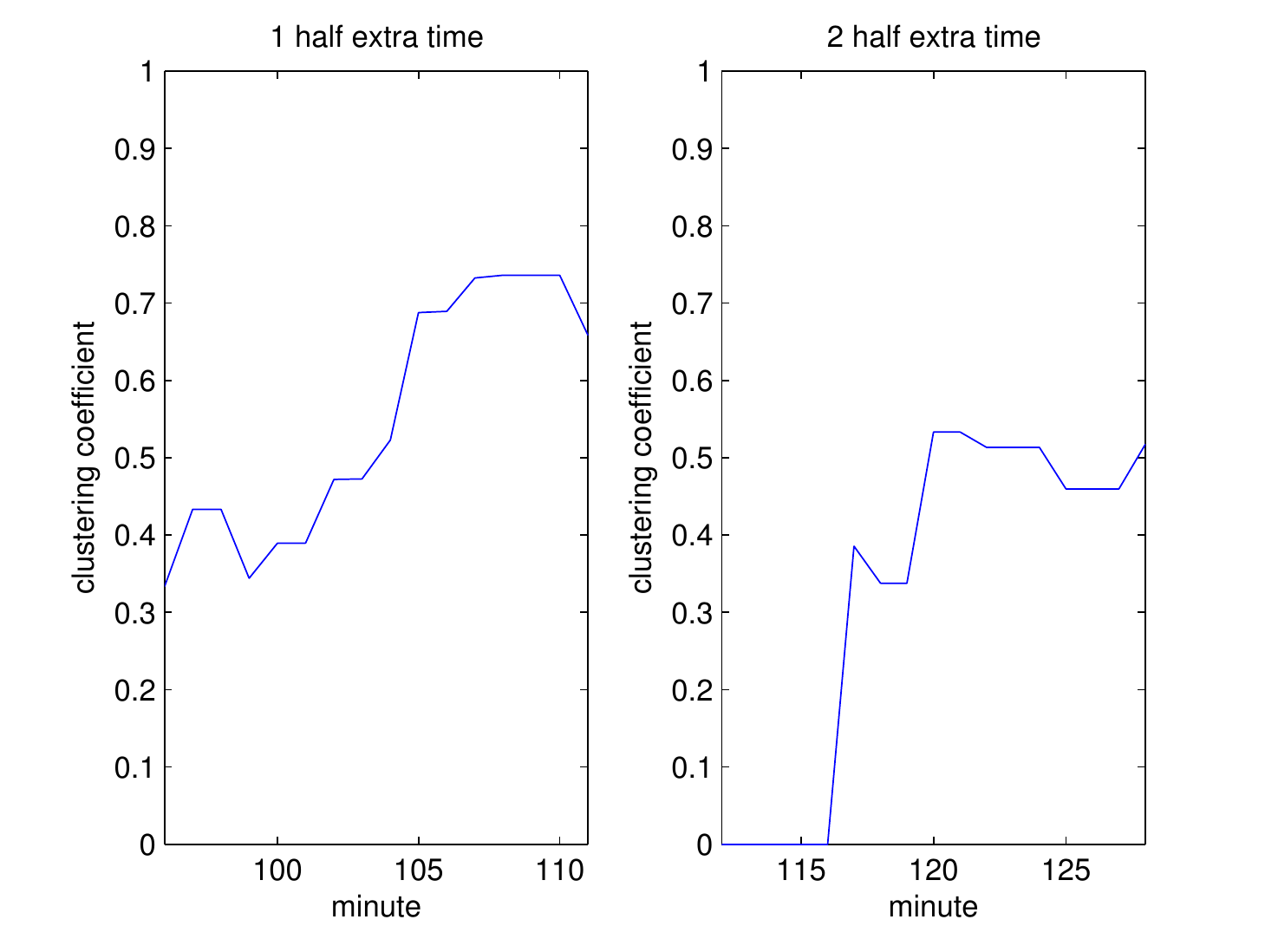}
\caption{Netherlands-Spain. Clustering rate in the extra time.
\label{fig:HvsS-cluster_rate-ET}}
\end{center}
\end{figure}

\begin{figure*}[!ht]
\begin{center}
\includegraphics[scale=0.5]{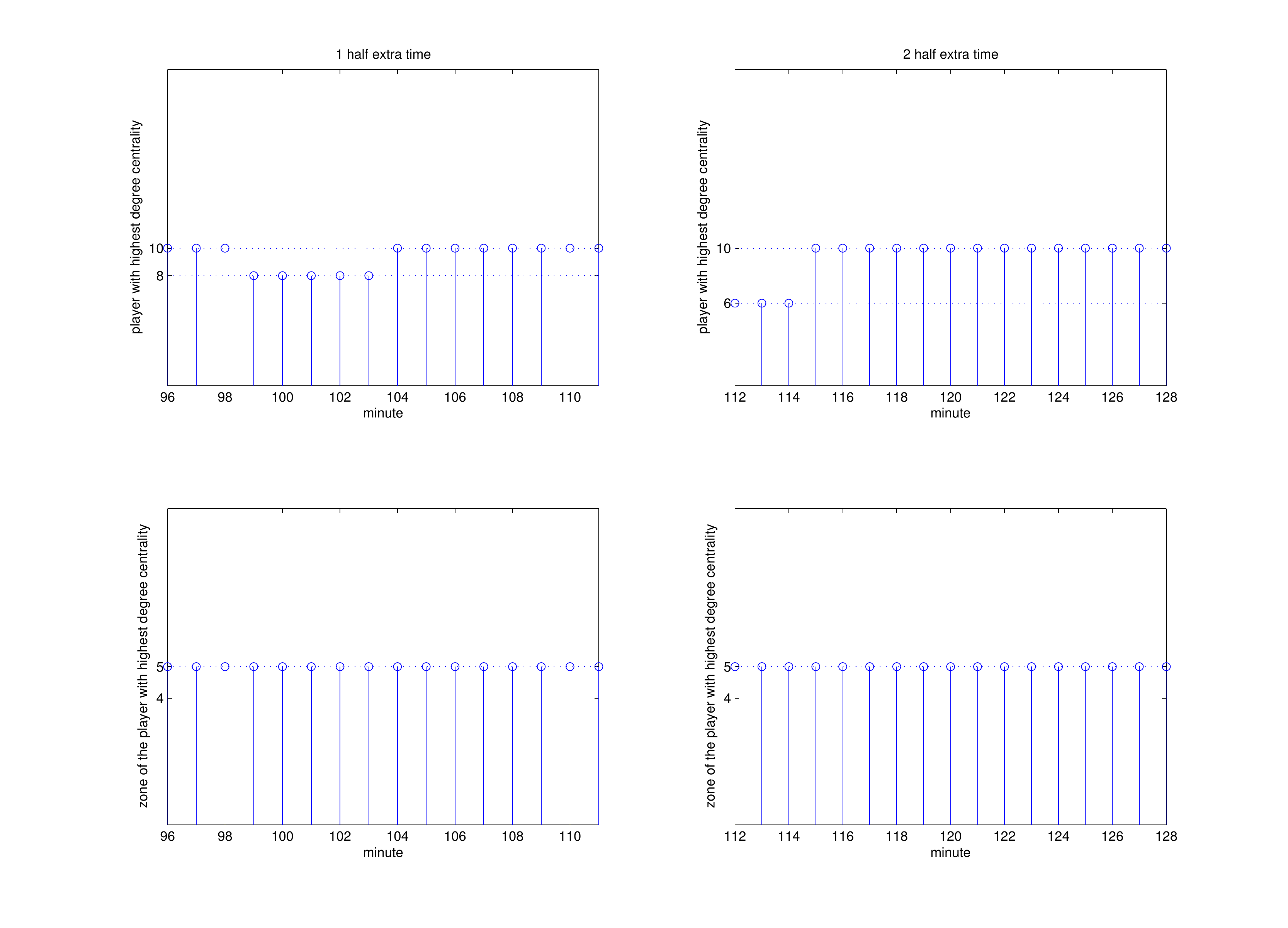}
\caption{Netherlands-Spain. Zone-based player centrality in the extra
  time. Please note that player number 6, is Andrés Iniesta, the one
  that scored the goal against the Netherlands (and which eventually
  led to victory).
\label{fig:HvsS-player_centrality-ET}}
\end{center}
\end{figure*}

The great final took place against the Netherlands, a very talented team that featured one of the best players of the season in the midfield (Wesley Sneijder, surprisingly left out of the Ballon d'Or contest) and possibly the fastest and finest winger in the whole tournament (Arjen Robben). Indeed, the connection among these two players could have very well  tilted the balance in favor of Netherlands had not \'{I}ker Casillas provided an outstanding performance. Despite the flair available on the pitch, the Dutch team engaged however in a very rough playing, forever iconized in Nigel de Jong's ominous tackle on Xabi Alonso's chest\footnote{\url{http://l-g.me/deJong}}. To some extent the tactical objective was similar to that of Paraguay's --disrupt the smooth flow of the ball in the Spanish team--- yet the execution was much more questionable (and so was the condescending attitude of the referee towards it). The game settled for the most part in a very slow pace due to the number of interruptions (Figure \ref{fig:HvsS-pass_minute}), and sequences of passes were generally short, analogously to the game against Paraguay. Unlike the latter game, the Dutch team succeeded in disabling Xabi Alonso but the balance of the game followed a similar pattern in that the physical style of the Netherlands became less consistent as time passed. The midfield was eventually controlled by Xavi Hern\'andez and --despite two clear fast breaks by Robben mentioned before-- the ball advanced away from Spain's defensive zone. The average number of consecutive passes gained momentum at the end of the 2nd half and was rather stable during the 1st part of the extra time. Not surprisingly, the pace of the game decreased steadily during the extra time (Figure \ref{fig:HvsS-pass_minute-ET}), due to both the exhaustion of the players and the team strategy shifting to a safer mode of holding possession in order to avoid conceding a late goal and trying to exploit a more direct approach when possible. Indeed, the clustering coefficient during the 2nd part of the extra time is much lower than in remaining games (Figure \ref{fig:HvsS-cluster_rate-ET}), which is consistent with a less structured development of the game and larger distances between team lines. The whole extra time witnessed the emergence of Cesc F\`abregas (Figure \ref{fig:HvsS-player_centrality-ET}), coming from the bench to dominate the midfield, providing eventually the assist that resulted in Andr\'es Iniesta's decisive goal.

\begin{figure}[!t]
\begin{center}
\includegraphics[scale=0.60]{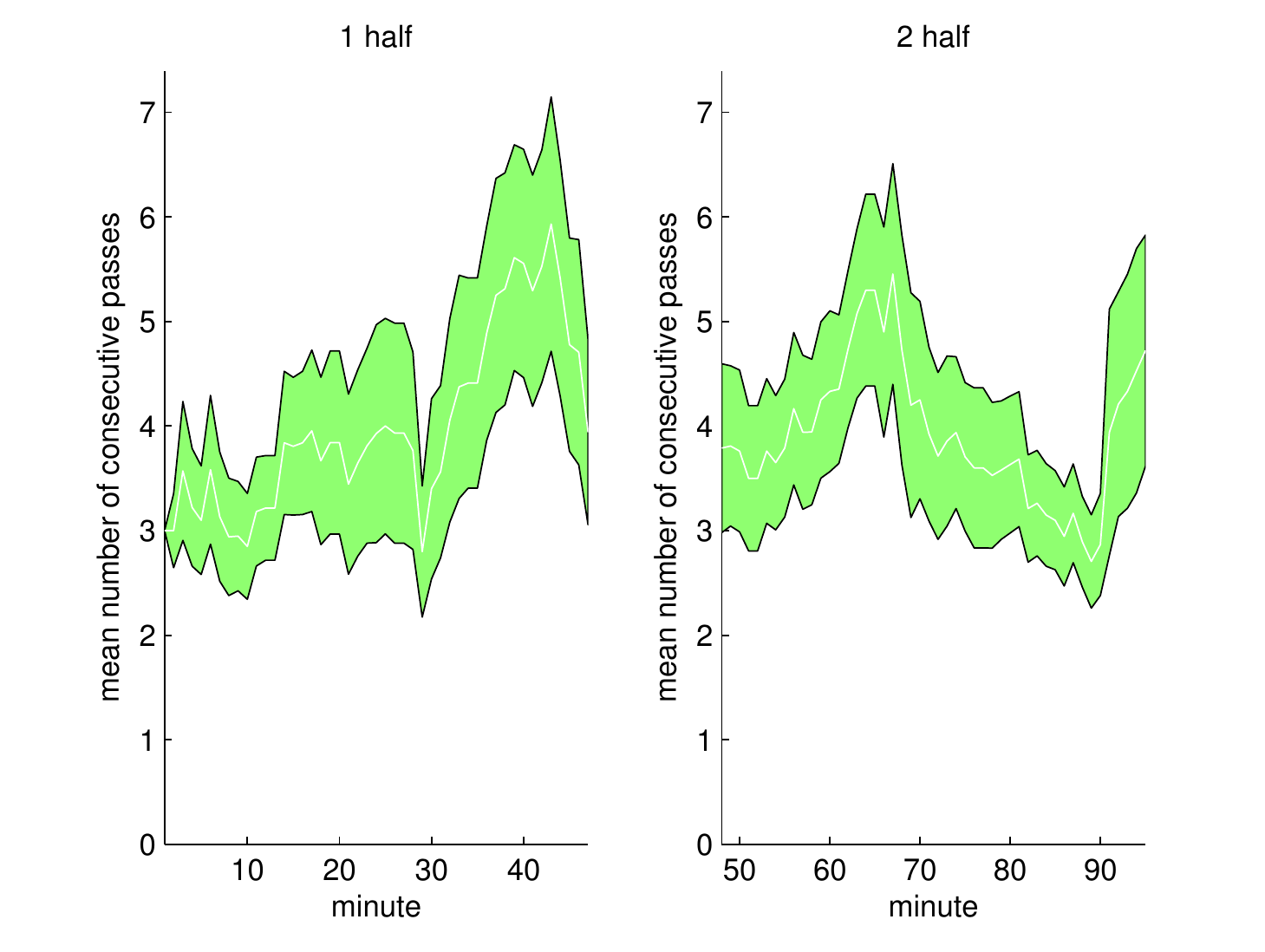}
\caption{Netherlands-Spain. Average number of consecutive passes.
\label{fig:HvsS-avg_pass_length}}
\end{center}
\end{figure}
\begin{figure}[!t]
\begin{center}
\includegraphics[scale=0.60]{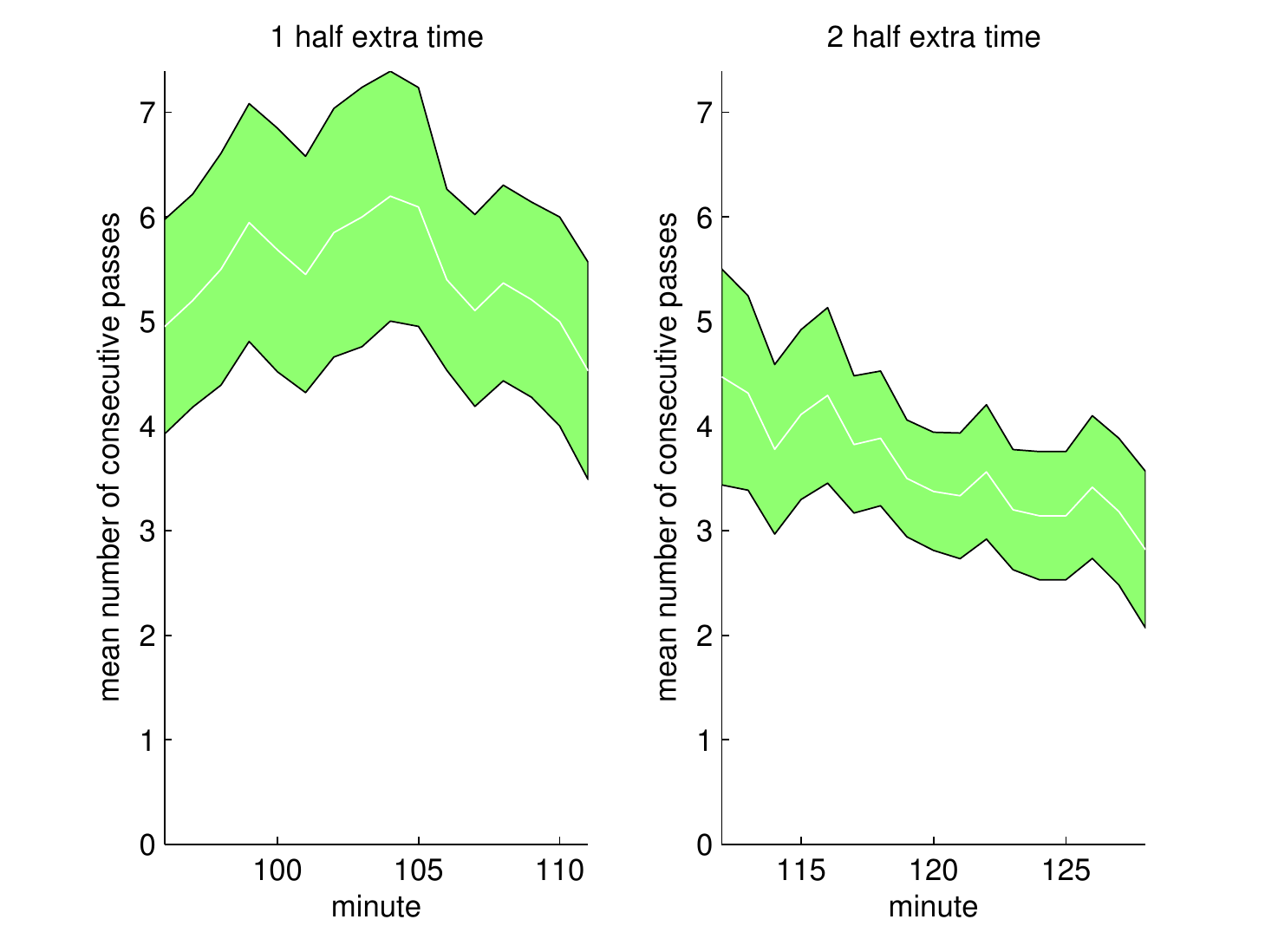}
\caption{Netherlands-Spain. Average number of consecutive passes in the extra time.
\label{fig:HvsS-avg_pass_length-ET}}
\end{center}
\end{figure}
\begin{figure}[!t]
\begin{center}
\includegraphics[scale=0.60]{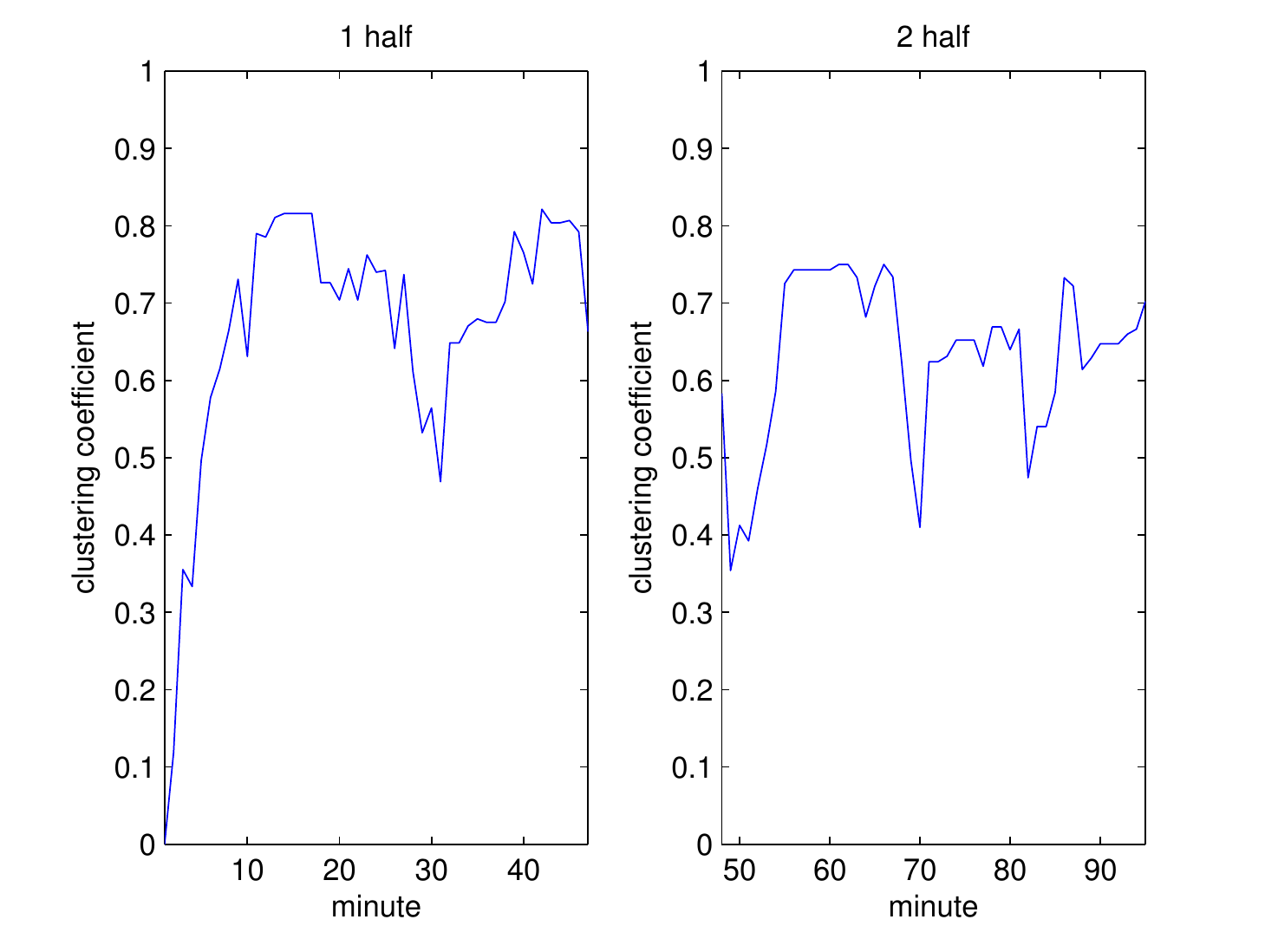}
\caption{Netherlands-Spain. Clustering rate.
\label{fig:HvsS-cluster_rate}}
\end{center}
\end{figure}
\begin{figure*}[!ht]
\begin{center}
\includegraphics[scale=0.5]{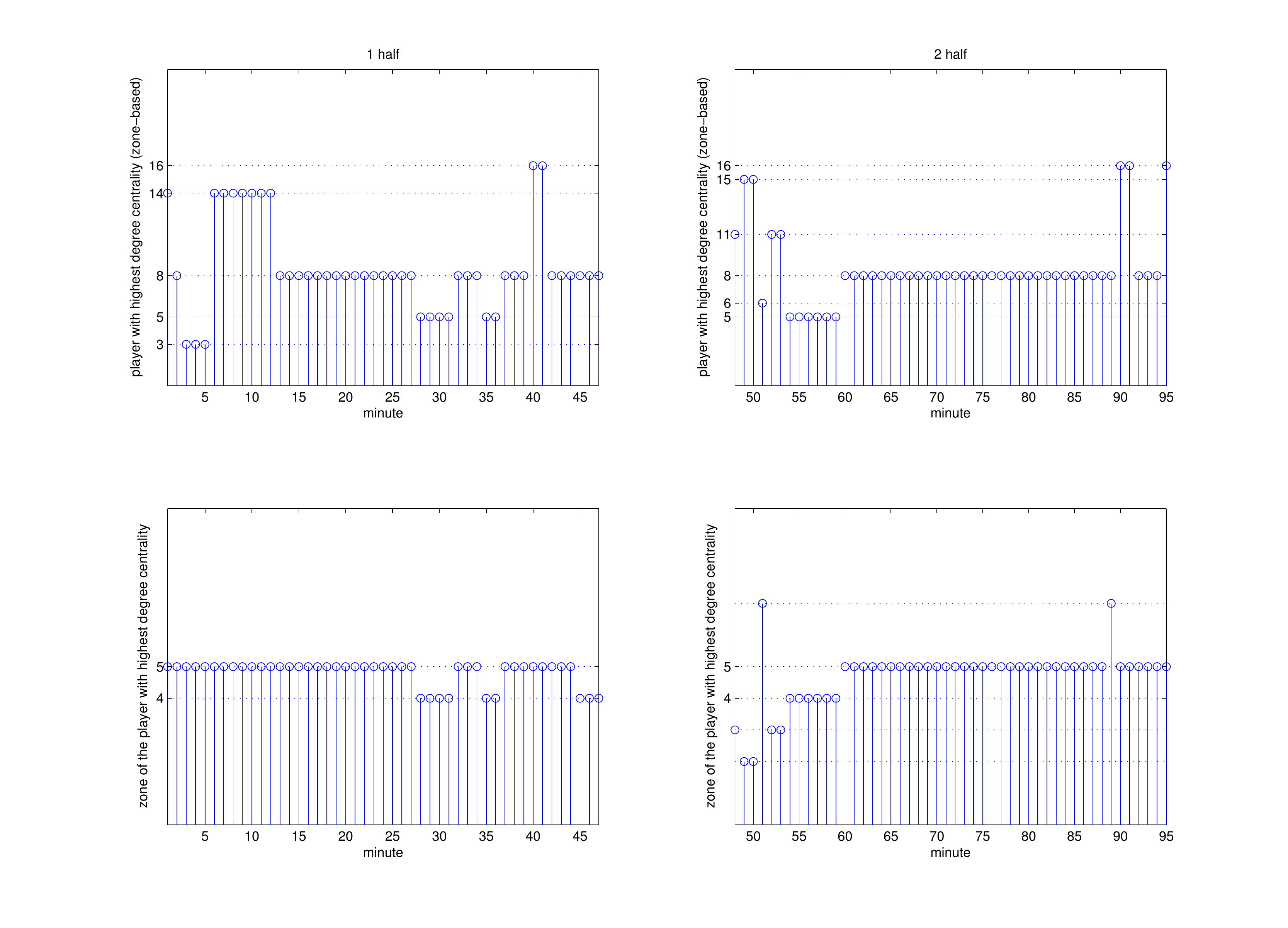}
\caption{Netherlands-Spain. Zone-based player centrality.
\label{fig:HvsS-player_centrality}}
\end{center}
\end{figure*}

\section{Conclusions and Future Work}
\label{conclusion}
This work has attempted to use some simple graph and network metrics
to analyze the performance and playing style of the Spanish national
football team in the World Cup 2010. When analyzed from a temporal
perspective, global measures such as the number of consecutive passes
or the number of passes per minute provide a measure of the success of
the team in imposing its style (or alternatively the success of the
opposing team in disrupting the style of the Spanish team). A deeper
insight is obtained by observing the clustering coefficient with
captures the combinative nature of the ``tiki-taka'' style. While
prone to a kind of baroqueness in the sense that many passes are
often done in a short distance and might be expendable, it is evident
that it has prime value as a defensive strategy, by depriving the
opponent of ball possession, one of the factors that is determinant in
the game outcome \citep{lago2007determinants,bate1988football}. Even in
the games in which the performance of the Spanish team has been deemed
worse by analysts and the general public, this combinativeness imprint
globally remains, with marked valleys in the aftermath of scoring a
goal (and the subsequent push of the trailing team) and sporadically
during the final, a special game in which emotions and passion often
led to lower precision and less elaborated game-play.

There is of course much work to be done. From the pure methodological
point of view, it is of crucial importance to keep analyzing other
network measures such as node and edge betweenness, which naturally
capture the hubs and essential associations in the distribution of the
ball (let us note en passant here as an example that the role of
Sergio Busquets was initially much criticized by the press and a large
part of the public due to his lower creativeness; however it is clear
that he was instrumental in providing balance to the midfield;
uncovering his role in the circuit of distribution of the ball by
means of an objective network metric would be of foremost
interest). Other measures such as eigenvector centrality can be very
valuable for this purpose. This can have not just an explanatory
value, but also a predictive value in terms of identifying the weak
points of the circuit that may be targeted by the opponent. In
addition to these methodological issues, it will be very interesting
to deploy this kind of analysis on further data both at club level
(e.g., to identify similarities and dissimilarities between team an
player performance in their clubs and in the national team), and at
the national team level such as in the upcoming Euro 2012 in Poland
and Ukraine or --looking beyond-- the World Cup 2014 of Brazil.

\section*{Acknowledgements}

This work has been supported in part by the CEI BioTIC GENIL (CEB09-0010) MICINN CEI Program (PYR-2010-13) project,
the Andalusian Regional Government P08-TIC-03903, P08-TIC-03928, and TIC-6083
projects, and MICINN TIN2008-05941. We would like to acknowledge the support of recently awarded
project TIN2011-28627-C04-02 and TIN2011-28627-C04-01.

\appendix
\section{Appendix. Spain's Squad Numbers}

\begin{table}[htbp]
\begin{tabular}{||ll|ll|ll||}
\hline
\hline
\# &~~& Name &~~& Position&~~\\
\hline
1	&&\'Iker Casillas 	&&Goalkeeper&\\
2	&&Ra\'ul Albiol	   && Defender&\\
3	&&Gerard Piqu\'e	  &&  Defender&\\
4	&&Carlos Marchena&&	Defender&\\
5	&&Carles Puyol	  &&  Defender&\\
6	&&Andr\'es Iniesta	&&Midfielder&\\
7	&&David Villa	   && Forward&\\
8	&&Xavi Hern\'andez	&&Midfielder&\\
9	&&Fernando Torres&&	Forward&\\
10	&&Cesc F\`abregas	&&    Midfielder&\\
11	&&Joan Capdevila&&	Defender&\\
12	&&V\'{\i}ctor Vald\'es	&&    Goalkeeper&\\
13	&&Juan Mata	    &&    Forward&\\
14	&&Xabi Alonso	  &&  Midfielder&\\
15	&&Sergio Ramos	 &&   Defender&\\
16	&&Sergio Busquets&&	Midfielder&\\
17	&&\'Alvaro Arbeloa&&	Defender&\\
18	&&Pedro Rodr\'{\i}guez&&	Forward&\\
19	&&Fernando Llorente&&	Forward&\\
20	&&Javi Mart\'{\i}nez	&&    Midfielder&\\
21	&&David Silva	  &&  Midfielder&\\
22	&&Jes\'us Navas	  &&  Forward&\\
23	&&Jos\'e Manuel Reina&& 	Goalkeeper&\\
\hline
\hline
\end{tabular}
\end{table}

\footnotesize
\bibliographystyle{apalike}
\bibliography{pases-mundial}

\begin{thebibliography}{}

\bibitem[Bate, 1988]{bate1988football}
Bate, R. (1988).
\newblock Football chance: tactics and strategy.
\newblock {\em Science and football}, pages 293--301.

\bibitem[Bundio and Conde, 2009]{bundio2009analisis}
Bundio, J. and Conde, M. (2009).
\newblock An{\'a}lisis del desempe{\~n}o deportivo durante la eurocopa 2004 a
  partir del an{\'a}lisis de redes sociales.
\newblock {\em REDES-Revista hispana para el an{\'a}lisis de redes sociales},
  13(0).

\bibitem[Dowie, 1982]{dowie1982spain}
Dowie, J. (1982).
\newblock Why {S}pain should win the world cup.
\newblock {\em New Scientist}, 94(10):693--695.

\bibitem[Heuer et~al., 2010]{heuer2010soccer}
Heuer, A., Mueller, C., and Rubner, O. (2010).
\newblock Soccer: Is scoring goals a predictable poissonian process?
\newblock {\em EPL (Europhysics Letters)}, 89:38007.

\bibitem[Heuer and Rubner, 2009]{springerlink:10.1140/epjb/e2009-00024-8}
Heuer, A. and Rubner, O. (2009).
\newblock Fitness, chance, and myths: an objective view on soccer results.
\newblock {\em The European Physical Journal B - Condensed Matter and Complex
  Systems}, 67:445--458.
\newblock 10.1140/epjb/e2009-00024-8.

\bibitem[Lago and Mart{\'\i}n, 2007]{lago2007determinants}
Lago, C. and Mart{\'\i}n, R. (2007).
\newblock Determinants of possession of the ball in soccer.
\newblock {\em Journal of Sports Sciences}, 25(9):969--974.

\bibitem[Onody and de~Castro, 2004]{onody2004complex}
Onody, R. and de~Castro, P. (2004).
\newblock Complex network study of brazilian soccer players.
\newblock {\em Physical Review E}, 70(3):037103.

\end{thebibliography}

\end{document}